\documentclass[aps,pra,eqsecnum,superscriptaddress,showpacs,floatfix,nofootinbib]{revtex4-1}
\usepackage{graphicx}
\usepackage{bm}
\usepackage{dsfont}
\usepackage{amsmath,amsfonts,amssymb}
\usepackage{float}
\usepackage{hyperref}
\usepackage{amsmath}
\usepackage{amsfonts}
\usepackage{amssymb}

\hypersetup{colorlinks=true,linkcolor=blue,citecolor=blue,urlcolor=blue}

\begin{document}

\title{Non-equilibrium dynamics of an ultracold Bose gas under a multi-pulsed quantum quench in interaction}

\author{Lei Chen}
\affiliation{Shenyang National Laboratory for Materials Science,
Institute of Metal Research, Chinese Academy of Sciences, Wenhua
Road 72, Shenyang 110016, China}

\author{Zhidong Zhang}
\affiliation{Shenyang National Laboratory for Materials Science,
Institute of Metal Research, Chinese Academy of Sciences, Wenhua
Road 72, Shenyang 110016, China}

\author{Zhaoxin Liang}
\email{Corresponding author: zhxliang@imr.ac.cn}
\affiliation{Shenyang National Laboratory for Materials Science,
Institute of Metal Research, Chinese Academy of Sciences, Wenhua
Road 72, Shenyang 110016, China}

\begin{abstract}
We investigate the nonequilibrium dynamical properties of a weakly-interacting
Bose gas at zero temperature under the multi-pulsed quantum quench
in interaction by calculating one-body, two-body correlation functions
and Tan's contact of the model system. The multi-pulsed quench is represented
as follows: first suddenly quenching the interatomic interaction
from $g_{i}$ to $g_{f}$ at time $t=0$, holding time $t$, and then
suddenly quenching interaction from $g_{f}$ back to $g_{i}$, holding
the time $t$ sequence $n$ times. In particular, two typical kinds
of quenching parameters are chosen, corresponding to $\left(g_{i}/g_{f}>1\right)$
and $\left(g_{i}/g_{f}<1\right)$ respectively. We find that the more
the quenching times of $n$ are, the more the excitations are excited,
which suggests that the multi-pulsed QQ is more powerful
way of studying the non-equilibrium dynamics of many-body quantum
system than the `one-off' quantum quench. Finally, we discuss the
ultra-short-range properties of the two-body correlation function
after the $n$th quenching, which can be used to probe the `Tan'scontact'
in experiments. All our calculations can be tested in current cold
atom experiments.
\end{abstract}

\maketitle

\section{Introduction}

Quantum quench (QQ) \cite{DynamicalRev.}, referred to as how a prepared
state based on the initial Hamiltonian of $H_{i}$ can evolve with
another Hamiltonian of $H_{f}$, provides a powerful tool for investigating
nonequilibrium dynamics of quantum many-body systems. Efforts
along this line have been driven, on the one hand, by the remarkable
progress in series of recent experiments \cite{Higher-resolution1,Higher-resolution2,ex4}
with ultracold atomic gases, on the other hand, by the desire to understand
the basic questions in nonequilibrium physics, ranging from thermalization
and equilibration \cite{DynamicalRev.,thermal1,thermal2,thermal3,thermal4,thermal5,thermal6}
and their relation to integrability \cite{int1,int2,int3,int4,int5},
to the introduction of new concept of dynamical phase transition \cite{phasetran1,phasetran2,phasetran3,tran4,tran5,tran6,thermal6}.
It can be said, in particular, that quenching ultracold atomic gases
has becoming the forefront for studying the nonequilibrium physics.

Up to now, typical scenarios for a QQ consist of a sudden change in
interatomic interaction due to Feshbach resonance or in the strength
of confining potential \cite{changeEpara1,changeEpara2Bragg,changeEpara3insitu,changeEpara4ToF,changeEpara5insitu,changeEpara6,changeEpara7},
characterized by the transition from $H_{i}$ to $H_{f}$ happening
over a time scale shorter than any other time scale in the problem.
The key ingredient of a QQ is that the final state arrived after a QQ
has more excitations than the corresponding equilibrium state, which
in turn measures the abilities of QQ driving the model system out
of equilibrium. Based on such a understanding, so far, most of the
previous theoretical studies \cite{Q1,Q2,Q3} have focused on the
`one-off' QQ defined by a QQ happens only once and subsequently
investigated how the key quantities of the model system, typical of
one-body or two-body correlation functions, can relax. However, the
`one-off' QQ sets a fundamental limit for the ability of driving the
model system out of equilibrium, which can be best understood in terms
of the Loschmidt echo (LE) $L(t)=|\langle\Psi_{0}(t)|\Psi(t)\rangle|^{2}$.
Here, the LE $L\left(t\right)$ is defined as the overlap of two wave
functions of $|\Psi_{0}(t)\rangle$ and $|\Psi(t)\rangle$ evolved
from the same initial state, but with different Hamiltonian $H_{i}$
and $H_{f}$. Remarkably, Refs. \cite{LE1,LE2} have proved that the
$L_{sq}\left(t\rightarrow\infty\right)$ under the sudden quench is
square of the adiabatic counterpart $L_{ad}\left(t\rightarrow\infty\right)$,
which is hold in general. This gives the maximum value of a `one-off'
QQ driving system out of equilibrium to $L_{sq}(t)=L^2_{ad}(t)$. Therefore,
how to further improve the power of a QQ inducing the out-of-equilibrium
dynamics, which is of fundamental interest to studying nonequilibrium physics
in a regime not accessible to a `one-off' QQ, has becoming a real challenge.

Here, we propose and analyze a novel kind of QQ denoted
to the multi-pulsed quench, wherein the Hamiltonian is quenched many times, in order to further enhance the out-of-equilibrium
dynamics. As shown in Fig. \ref{Fig1}. the typical protocol of a multi-pulsed QQ studied in this work
consists of as follows:
at time $t=0$, suddenly quenching the interatomic interaction from
$g_{i}$ to $g_{f}$, holding time $t$, and then suddenly quenching
interaction from $g_{f}$ back to $g_{i}$, holding the time $t$
sequence $n$ times. The power of our approach can be illustrated
in terms of general physical arguments based on the LE $L_{mp}(t)$
corresponding to a multi-pulsed QQ.
It can be proved that $L_{mp}(2nt)$ after $n$-time multi-pulsed quench
is the $2n$-th power of the adiabatic counterpart $L_{ad}\left(t\right)$,
{\it i.e.} $L_{mp}(2nt)=L_{sq}^{n}\left(t\right)=L_{ad}^{2n}\left(t\right)$
\cite{LE1,LE2}. The fact that the exponent is enhanced by a factor
of $n$ compared to the `one-off' quench shows that more far away
equilibrium regime can be reached. Moreover, Our proposal hints at the possibility
to induce the non-equilibrium dynamics in a highly controllable way.

In this work, motivated by the recent development of quenching ultra-cold
atom systems in both the experimental and the theoretical sides, we have launched systematic
studies on the effects of the multi-pulsed QQ on two kinds of
correlation functions of a three-dimensional (3D) ultracold Bose gas, {\it i.e.} one-body,
two-body correlation functions and Tan's contact \cite{Contact1,Contact2Bragg,Contact3,Contact4,Contact5,Contact6,Contact7}
of the model system respectively. The reasons are two-fold: first,
these two (one- and two-body) correlation functions provide direct insight
into the dynamical properties of both equilibrium and nonequilibrium
quantum many-body system; second, the one-body correlation function
is directly connected with the noncondensated fraction which can be
measured in time-of-flight experiments \cite{changeEpara4ToF}.
The two-body correlation function denotes the correlation between two
particles in different spatial positions at the same time and can
be detected using Bragg spectroscopy \cite{changeEpara2Bragg,Contact2Bragg,Bragg},
noise correlations \cite{NoiseCorrelation1,NoiseCorrelation2}, and
even by in situ measurements \cite{changeEpara3insitu,changeEpara5insitu}.
Moreover, we also discuss the ultra-short-range properties of the two-body
correlation function, which is related to the internal energy via
the `Tan's contact' proposed by Tan \cite{Contact1} in the system
of a two-component Fermi gas interacting with a contact interaction.
In more details, we use the time-dependent Bogoliubov approximation
to study the non-equilibrium dynamics of a 3D Bose gas after a multi-pulsed
quench. In particular, by changing quench times $n$, we try to
collect information about how correlation functions evolve with time.
We have found that the model system can produce more elementary excitations
with increasing the quench times $n$ as expected. Although the two-body correlation
function oscillates fast in short-time-range, these two kinds of correlation
functions tend to a constant value eventually even we
increase the quenching times to $12$. Finally, we also find that the Tan's contact changes slightly
with time, which almost is independent of the quench times.

\section{3D Bose gas under a multi-pulsed QQ in interaction}

\label{sec:Model}

\begin{figure}[!tbh]
\begin{center}
\rotatebox{0}{\resizebox *{8.5cm}{4.5cm} {\includegraphics
{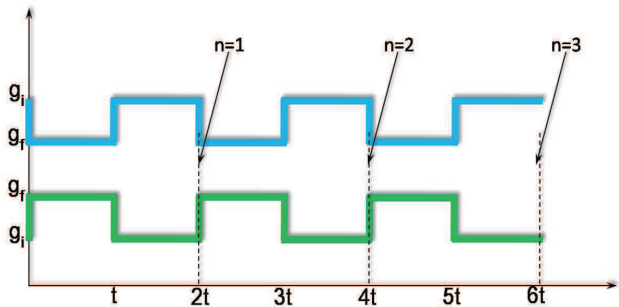}}}
\end{center}
\protect\caption{(Color online) Schematics of the multi-pulsed quantum quench where only three
repeating sequences are visible. The upper (blue) one refers to quenching
from $g_{i}$ to $g_{f}$ with the ratio $g_{i}/g_{f}>1$;
The lower (green) one denotes the opposite case with the ratio $g_{i}/g_{f}<1$.
The arrows show the time instant corresponding to
the $n$th order quench.}
\label{Fig1}
\end{figure}

The model system considered in this work is composed of a 3D Bose
gas of $N$ ultracold bosonic atoms. In addition, the interatomic
interaction is assumed to be weak and can be well described by the contact
interaction. In such, the many-body Hamiltonian under a multi-pulsed QQ in the interaction reads as follows
\begin{equation}
H=\sum_{\mathbf{k}}\left(\epsilon_{\mathbf{k}}-\mu\right)a_{\mathbf{k}}^{\dagger}a_{\mathbf{k}}+\frac{g\left(t\right)}{2V}\sum_{\mathbf{k_{1}},\mathbf{k_{2}},\mathbf{q}}a_{\mathbf{k_{1}}}^{\dagger}a_{-\mathbf{k_{1}}+\mathbf{q}}^{\dagger}a_{\mathbf{k_{2}}}a_{-\mathbf{k_{2}}+\mathbf{q}},\label{eq:H}
\end{equation}
with $a_{\mathbf{k}}^{\dagger}$ and $a_{\mathbf{k}}$ being the bosonic
creation and annihilation operators, $\epsilon_{\mathbf{k}}=\hbar^{2}k^{2}/\left(2m\right)$
labeling the single-particle dispersion relation and $\mu$ and $V$
being the chemical potential and the volume of the system respectively.
The $g\left(t\right)$ in Hamiltonian (\ref{eq:H}) describes the
multi-pulsed quench protocol. To be specific,
as shown in Fig. \ref{Fig1}, we consider the case: (i) the system is initially
prepared at the ground state $|\Psi_{0}(t)\rangle$ of Hamiltonian
(\ref{eq:H}) with $g(t)=g_{i}$ labeled by $H_{i}$; (ii) then, at $t=0$, the
interaction strength is suddenly switched to $g(t)=g_{f}$ such that
the time evolution from $t>0$ is governed by the finial Hamiltonian
(\ref{eq:H}) of $H_{f}$ ; (iii) after holding the time $t$, the interaction
suddenly quenches form $g_{f}$ back to $g_{i}$, holding the time
$t$ sequence $n$ times.

We focus on the regime of weak interatomic interaction, in which Hamiltonian
(\ref{eq:H}) can be well described by the standard Bogoliubov approximation.
As a standard fashion, we proceed to transform the Hamiltonian (\ref{eq:H})
into the effective Hamiltonian: the zeroth-order term is found by
substituting all creation and annihilation operator by $\sqrt{N_{0}}$
with $N_{0}=\left\langle a_{\mathbf{k}=0}\right\rangle ^{2}$ being
the number of condensed atoms; the first-order term is found by
keeping only one creation or annihilation operators in every term
of the Hamiltonian; the quartic
or more higher term can be obtained following the same logic, but
we retain only to the second-order term because of the terms higher than two
doing little contribution to our question were omitted. Collecting
all terms calculated above, we arrive at the effective Hamiltonian:
\begin{eqnarray}
H^{\mathrm{eff}} & = & -\frac{1}{2}g\left(t\right)n_{0}N_{0}+\sum_{\mathbf{k}\ne0}\left(\epsilon_{\mathbf{k}}+g\left(t\right)n_{0}\right)a_{\mathbf{k}}^{\dagger}a_{\mathbf{k}}\nonumber \\
 &  & +\frac{1}{2}g\left(t\right)n_{0}\sum_{\mathbf{k}\ne0}\left(a_{\mathbf{k}}a_{-\mathbf{k}}+a_{-\mathbf{k}}^{\dagger}a_{\mathbf{k}}^{\dagger}\right),\label{eq:EffecH}
\end{eqnarray}
with $n_{0}=N_{0}/V$ being the condensed density.
Next, the effective Hamiltonian (\ref{eq:EffecH}) can be diagonalized
by a standard Bogoliubov variational ansatz, where we write the
bosonic creation $a_{\mathbf{k}}^{\dagger}$ and annihilation $a_{\mathbf{k}}$
operators in new defined operators $b_{\mathbf{k}}^{\dagger}$ and
$b_{\mathbf{k}}$,
\begin{eqnarray}
a_{\mathbf{k}\ne0}\left(t\right) & = & u_{\mathbf{k}}\left(t\right)b_{\mathbf{k}}+v_{\mathbf{k}}^{*}\left(t\right)b_{-\mathbf{k}}^{\dagger},\nonumber \\
a_{-\mathbf{k}\ne0}^{\dagger}\left(t\right) & = & v_{\mathbf{k}}\left(t\right)b_{\mathbf{k}}+u_{\mathbf{k}}^{*}\left(t\right)b_{-\mathbf{k}}^{\dagger},\label{eq:creannioperators}
\end{eqnarray}
with $b_{\mathbf{k}}^{\dagger}$ and $b_{\mathbf{k}}$ denoting the
bosonic creation and annihilation operators for noncondensed atoms,
respectively. These two operators have no time dependence and always
are treated as small quantities. To assure operators $b_{\mathbf{k}}^{\dagger}$
and $b_{\mathbf{k}}$ still comply with the standard commutation relations
for bosonic creation and annihilation operators, we have double-checked
the relation $\left|u_{\mathbf{k}}\left(t\right)\right|^{2}-\left|v_{\mathbf{k}}\left(t\right)\right|^{2}=1$,
which is always satisfied. Substituting Eq. (\ref{eq:creannioperators})
into Eq. (\ref{eq:EffecH}), we can obtain the diagonalized Hamiltonian
expressed by operators $b_{\mathbf{k}}^{\dagger}$ and $b_{\mathbf{k}}$. Furthermore, at time
$t=0$, we find that coherence factors $u_{\mathbf{k}}$
and $v_{\mathbf{k}}$ must be solutions of the following equations:
\begin{eqnarray}
gn_{0}\left(u_{\mathbf{k}}^{2}+v_{\mathbf{k}}^{2}\right)+2\left(\epsilon_{\mathbf{k}}+gn_{0}\right)u_{\mathbf{k}}v_{\mathbf{k}} & = & 0,\nonumber \\
\left(\epsilon_{\mathbf{k}}+gn_{0}\right)\left(\left|u_{\mathbf{k}}\right|^{2}+\left|v_{\mathbf{k}}\right|^{2}\right)+gn_{0}\left(v_{\mathbf{k}}^{*}u_{\mathbf{k}}+u_{\mathbf{k}}^{*}v_{\mathbf{k}}\right) & = & \hbar\omega_{\mathbf{k}}.\nonumber \\
\end{eqnarray}
Combined with the normalization $\left|u_{\mathbf{k}}\right|^{2}-\left|v_{\mathbf{k}}\right|^{2}=1$,
we can easily find the solutions as follows
\begin{eqnarray}
u_{\mathbf{k}}\left(t=0\right) & = & \sqrt{\left[\left(\epsilon_{\mathbf{k}}+g_{i}n_{0}\right)/E_{\mathbf{k}}^{i}+1\right]/2},\nonumber \\
v_{\mathbf{k}}\left(t=0\right) & = & -\sqrt{\left[\left(\epsilon_{\mathbf{k}}+g_{i}n_{0}\right)/E_{\mathbf{k}}^{i}-1\right]/2}\label{eq:cohfact}
\end{eqnarray}
with $\hbar\omega_{\mathbf{k}}=E_{\mathbf{k}}^{i}=\sqrt{\epsilon_{\mathbf{k}}\left(\epsilon_{\mathbf{k}}+2g_{i}n_{0}\right)}$.
Finally, the diagonalized Hamiltonian reads as:
\begin{equation}
H^{\mathrm{eff}}=-\frac{1}{2}gn_{0}N_{0}+\frac{1}{2}\sum_{\mathbf{k}\ne0}\left[\hbar\omega_{\mathbf{k}}-\left(\epsilon_{\mathbf{k}}+gn_{0}\right)\right]+\sum_{\mathbf{k}\ne0}\hbar\omega_{\mathbf{k}}b_{\mathbf{k}}^{\dagger}b_{\mathbf{k}}.\label{eq:EffecHB}
\end{equation}
\begin{figure}[!tbh]
\begin{center}
\rotatebox{0}{\resizebox *{8.5cm}{4.5cm} {\includegraphics
{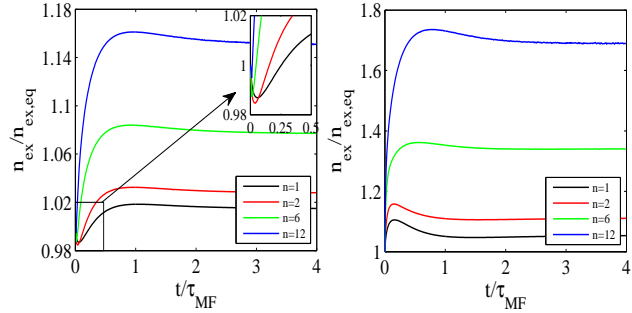}}}
\end{center}
\caption{(Color online) Quasimomentum distribution  (one-body matrix) via the time $t$ with different
values of $n$ in the multi-pulsed quantum quench. The left panel refers to quenching to a
smaller interaction with $g_{i}/g_{f}=1.1$, and the right panel denotes the opposite case
with $g_{i}/g_{f}=0.8$. The fraction of noncondensed
atoms in both cases are all normalized to the initial excitation fraction.
The characteristic relaxation time is set by $\tau_{\mathrm{MF}}=\hbar/g_{f}n_{0}$. The quenching times are selected
to $n=1,2,6,12$.}
\label{Fig2}
\end{figure}
\begin{figure}[!tbh]
\begin{center}
\rotatebox{0}{\resizebox *{8.5cm}{4.5cm} {\includegraphics
{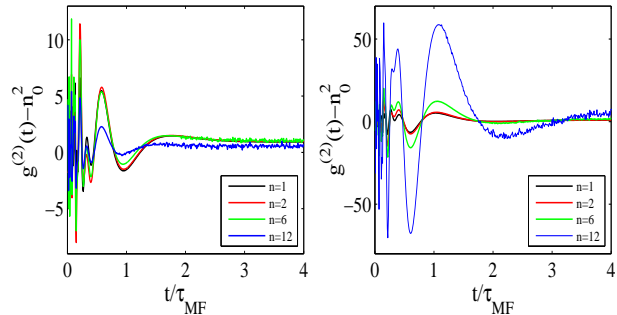}}}
\end{center}
\protect\caption{(Color online) Plotted are nonequilibrium dynamics of density-density
correlations (two-body correlation function) via the time $t$. Density-density correlations $g^{\left(2\right)}\left(t\right)-n_{0}^{2}$
is normalized to the asymptotic value at long times. Time and length scale are measured
in terms of the $\tau_{\mathrm{MF}}=\hbar/g_{f}n_{0}$
and  $\left(\zeta=\hbar/\sqrt{mg_{f}n_{0}}\right)$
respectively.  The dimensionless distance between two different
spatial positions is choosen to be $\delta/\zeta=4$. The quenching times in multi-pulsed quantum quench are selected to $n=1,2,6,12$.}
\label{Fig3}
\end{figure}

Both the one-body and the two-body correlation functions can be expressed
as a function of  $u_{\mathbf{k}}\left(t\right)$
and $v_{\mathbf{k}}\left(t\right)$. Hence how to get these two factors
is the core of issue. We need to work in the Heisenberg representation
and make use of the equations of motion for $a_{\mathbf{k}}\left(t\right)$ as follows,
\begin{equation}
i\hbar\partial_{t}a_{\mathbf{k}}\left(t\right)=\left[
a_{\mathbf{k}}\left(t\right), H^{\mathrm{eff}}\right]\label{eq:Eq.of motion}
\end{equation}
The solving process of $u_{\bf k}(t)$ and $v_{\bf k}(t)$ is straightforward but somewhat complicated, we only give a number of important steps.
Substituting Eq. (\ref{eq:creannioperators})
into Eq. (\ref{eq:Eq.of motion}), we can get differential equations
for $u_{\mathbf{k}}\left(t\right)$ and $v_{\mathbf{k}}\left(t\right)$:
\begin{equation}
i\hbar\partial_{t}\left(\begin{array}{c}
u_{\mathbf{k}}\\
v_{\mathbf{k}}
\end{array}\right)=\left(\begin{array}{cc}
\epsilon_{\mathbf{k}}+gn_{0} & gn_{0}\\
-gn_{0} & -\left(\epsilon_{\mathbf{k}}+gn_{0}\right)
\end{array}\right)\left(\begin{array}{c}
u_{\mathbf{k}}\\
v_{\mathbf{k}}
\end{array}\right)\label{eq:partialU,V}
\end{equation}
We can solve Eqs. (\ref{eq:partialU,V}) based on the initial conditions
$u_{\mathbf{k}}\left(t=0\right)$ and $v_{\mathbf{k}}\left(t=0\right)$,
reading,
\begin{equation}
\left(\begin{array}{c}
u_{\mathbf{k}}\left(t\right)\\
v_{\mathbf{k}}\left(t\right)
\end{array}\right)=U_{i\rightarrow f}\left(t\right)\left(\begin{array}{c}
u_{\mathbf{k}}\left(0\right)\\
v_{\mathbf{k}}\left(0\right)
\end{array}\right)\label{eq:Solutions}
\end{equation}
where the time evolution operator $U_{i\rightarrow f}\left(t\right)$ is defined as Eq. (\ref{eq:evolutionmatrix})
\begin{widetext}
\begin{equation}
U_{i\rightarrow f}\left(t\right)=\left(\begin{array}{cc}
\cos\left(E_{\mathbf{k}}^{f}t\right)-i\frac{\epsilon_{\mathbf{k}}+g_{f}n_{0}}{E_{\mathbf{k}}^{f}}\sin\left(E_{\mathbf{k}}^{f}t\right) & -i\frac{g_{f}n_{0}}{E_{\mathbf{k}}^{f}}\sin\left(E_{\mathbf{k}}^{f}t\right)\\
i\frac{g_{f}n_{0}}{E_{\mathbf{k}}^{f}}\sin\left(E_{\mathbf{k}}^{f}t\right) & \cos\left(E_{\mathbf{k}}^{f}t\right)+i\frac{\epsilon_{\mathbf{k}}+g_{f}n_{0}}{E_{\mathbf{k}}^{f}}\sin\left(E_{\mathbf{k}}^{f}t\right)
\end{array}\right).\label{eq:evolutionmatrix}
\end{equation}
\end{widetext}

In the derivation of the expressions of the time-dependent coherence
factors $u_{\mathbf{k}}\left(t\right)$ and $v_{\mathbf{k}}\left(t\right)$,
we limit ourselves into the regime where the time dependence of $n_{0}$
can be safely neglected as shown in Refs. \cite{Q1,n0-nNice}.
Moreover, the condition of $n_{ex}\ll n$ can be easily obtained in
the typical ultra-cold atomic experiments.

\section{Nonequilibrium dynamics of one- and two-body correlation functions}

\label{sec:results}

\begin{figure}[!tbh]
\begin{center}
\rotatebox{0}{\resizebox *{8.5cm}{6cm} {\includegraphics
{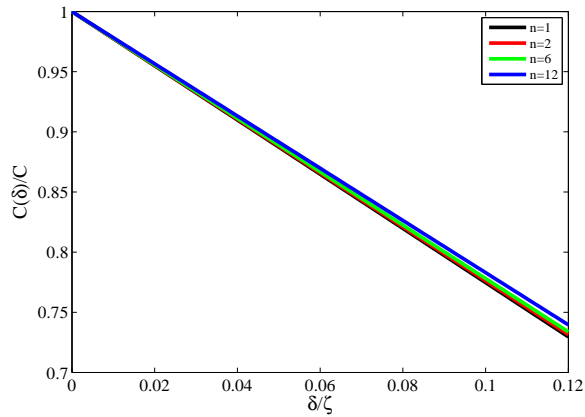}}}
\end{center}
\protect\caption{(Color online) Plotted are the `Tan's relation'  $\mathcal{C}\left(\delta\right)=\delta^{2}g^{\left(2\right)}\left(\delta\right)$ via $\delta$. At $t<0$,
the model system is noninteracting with $\mathcal{C}\left(\delta\right)=0$.
Immediately after the quench, the zero-distance correlations respond
instantaneously. Length and time scales are measured in terms of the
condensate healing length $\left(\zeta=\hbar/\sqrt{mg_{f}n_{0}}\right)$
and mean-field time $\tau_{\mathrm{MF}}=\hbar/g_{f}n_{0}$ in the final
state respectively.  The colored curves correspond to different quench times (bottom to top):
$n=1$ (black), $n=2$ (red), $n=6$ (green), and $n=12$ (blue).
The parameters read  $t/\tau_{\mathrm{MF}}=1$, $g_{i}/g_{f}=1.1$ and $a=0.5\zeta$.}
\label{Fig4}
\end{figure}

Using  $u_{\mathbf{k}}\left(t\right)$
and $v_{\mathbf{k}}\left(t\right)$, the multiple quench scenario (see in Fig. (\ref{Fig1})) can be expressed as follows. The Bogoliubov coefficients
of $u_{\mathbf{k}}^{\left(n\right)}\left(t\right)$
and $v_{\mathbf{k}}^{\left(n\right)}\left(t\right)$ at time $t$ after the $n$-th quench can
be determined as follows,
\begin{equation}
\left(\begin{array}{c}
u_{\mathbf{k}}^{\left(n\right)}\left(t\right)\\
v_{\mathbf{k}}^{\left(n\right)}\left(t\right)
\end{array}\right)=\left[U_{f\rightarrow i}\left(t\right)U_{i\rightarrow f}\left(t\right)\right]^{n}\left(\begin{array}{c}
u_{\mathbf{k}}\left(0\right)\\
v_{\mathbf{k}}\left(0\right)
\end{array}\right)\label{eq:bangbangprotocol}
\end{equation}
Then the one- and two-body correlation functions for $n$th quench
protocol, according to their definitions $n_{\mathrm{ex}}\left(t\right)=\sum_{\mathbf{k}\ne0}\left\langle a_{\mathbf{k}}^{\dagger}\left(t\right)a_{\mathbf{k}}\left(t\right)\right\rangle $
and $g^{\left(2\right)}\left(\mathbf{r}-\mathbf{r^{\prime}}\right)\left(t\right)=\sum_{\mathbf{q}}e^{i\mathbf{q}\cdot\left(\mathbf{r}-\mathbf{r^{\prime}}\right)}\left\langle \rho_{\mathbf{q}}\left(t\right)\rho_{-\mathbf{q}}\left(t\right)\right\rangle $
with $\rho_{\mathbf{q}}\left(t\right)=\sum_{\mathbf{k}}a_{\mathbf{k}+\mathbf{q}}^{\dagger}\left(t\right)a_{\mathbf{k}}\left(t\right)$, can be expressed as follows
\begin{equation}
n_{\mathrm{ex}}^{\left(n\right)}\left(t\right)=\sum_{\mathbf{k}}\left|v_{\mathbf{k}}^{\left(n\right)}\left(t\right)\right|^{2},\label{eq:onebody}
\end{equation}
and
\begin{eqnarray}
g^{\left(2\right)\left(n\right)}\left(\mathbf{\delta}\right)\left(t\right) & = & n_{0}^{2}+n_{0}\sum_{\mathbf{k}}e^{i\mathbf{k}\cdot\mathbf{\delta}}\left(u_{\mathbf{k}}^{\left(n\right)*}\left(t\right)v_{\mathbf{k}}^{\left(n\right)}\left(t\right)\right.\nonumber \\
 &  & \left.+u_{\mathbf{k}}^{\left(n\right)}\left(t\right)v_{\mathbf{k}}^{\left(n\right)*}\left(t\right)+2\left|v_{\mathbf{k}}^{\left(n\right)}\left(t\right)\right|^{2}\right)\label{eq:twobody}
\end{eqnarray}
with $\mathbf{\delta}=\mathbf{r}-\mathbf{r^{\prime}}$. We point out that the terms
quartic in $u_{\mathbf{k}}$'s and $v_{\mathbf{k}}$'s arising
from correlations between the noncondensed atoms are ignored because these correlations
become unimportant at long distances $\mathbf{\delta}>a\sim50$
nm; in contrast, the short-distance structure of the two-body correlation will
become very important. Therefore, we also derive the concrete expression
for Tan's contact as follows,
\begin{eqnarray}
\mathcal{C}\left(\mathbf{\delta}\right) & = & 16\pi^{2}\lim_{\mathbf{\delta}\rightarrow0}\mathbf{\delta}^{2}\left[\left(\sum_{\mathbf{k}}e^{-i\mathbf{k}\cdot\mathbf{\delta}}\left|v_{\mathbf{k}}^{\left(n\right)}\left(t\right)\right|^{2}\right)^{2}\right.\nonumber \\
 &  & \left.+\left|\sum_{\mathbf{k}}e^{-i\mathbf{k}\cdot\mathbf{\delta}}u_{\mathbf{k}}^{\left(n\right)}\left(t\right)v_{\mathbf{k}}^{\left(n\right)*}\left(t\right)\right|^{2}\right].\label{eq:contact}
\end{eqnarray}
Throughout this paper we consider two typical kinds of the multi-pulsed QQ: quench to a smaller interaction $g_{i}/g_{f}>1$ or quench
to a bigger interaction $g_{i}/g_{f}<1$. In both cases, the noncondensed
fractions, and the equal-time density-density correlation functions,
and the Tan's contacts all have analytical expressions, however, they
are a bit complicated, so we not list them here. In what follows,
we focus on the influence of the multi-pulsed QQ on two kinds of
correlation functions as shown in Figs. \ref{Fig2}, \ref{Fig3} and \ref{Fig4}, which
are plotted on the basis of their analytic expressions. In this end,
we devise following two scenarios: first, we refer to $n=1,2,6,12$
as the times of quench in the quench protocol and then study
how the correlation functions evolve with time after the
quench with the different values of $n$; second, we choose the parameter regimes of both $g_{i}/g_{f}>1$
and $g_{i}/g_{f}<1$ and consider how the detailed changes of interaction affect
the non-equilibrium dynamics of correlation functions.

In the first scenario, the results are plotted in Figs. \ref{Fig2}, \ref{Fig3} and \ref{Fig4}.
As shown in Fig. \ref{Fig2}, it is evident that the multi-pulsed QQ can
produce more noncondensed fraction with increasing the quench times of $n$
as expected, which suggests that more far away equilibrium regimes can
be achieved. Moreover, both the rapid relaxation of the one-body matrix
in Fig. \ref{Fig2} and two-body correlation in Fig. \ref{Fig3} suggests that, after the
3D Bose gas is brought out of equilibrium by a QQ, it can relaxes to a steady state on a time-scale within the
experimental reach. In Fig. \ref{Fig4}, the fact that the quench times of $n$
has the relative small effects on the Tan's relation can be understood
as follows: a QQ usually induces the low-energy fluctuations compared
to the energy scale which can contribute to Tan's relation.

In the second scenario, as shown in the left and right panels of Figs.
\ref{Fig2} and \ref{Fig3}, we compare the effects of the different choices of $g_{i}/g_{f}>1$
and $g_{i}/g_{f}<1$ on correlation functions. It's clear that the
more bigger is the final interaction $g_{f}$, the more excitations
a QQ can induce as shown in Fig. \ref{Fig2}. As for two-body correlation functions
in Fig. \ref{Fig3}, they will all develop the rapid oscillation matter in short-time
range, and evolve to a final equilibrium state as long as the time
is long enough. Moreover, the oscillate amplitude decrease with increasing the
quench times when the ratio $g_{i}/g_{f}>1$, and will increase when
the ratio $g_{i}/g_{f}<1$ (see Fig. \ref{Fig4})

\section{Conclusions} \label{sec:conclusions}

In summary, we first investigate the non-equilibrium dynamics of one-
and two-body correlation functions in a 3D homogeneous Bose gas at
zero temperature following the multi-pulsed QQ in interaction. Our
results of one- and two-body correlation functions show that the multi-pulsed
QQ can bring the model system far more away equilibrium regime than
the `one-off' QQ, which suggests that the multi-pulsed QQ is more powerful
way of studying the non-equilibrium dynamics of many-body quantum
system.

\section*{Acknowledgments}

We thanked Zhu Chen for helping on numerical calculations. This work
is supported by the NSF of China (Grant Nos. 11004200 and 11274315).


\begin{thebibliography}{99}

\bibitem{DynamicalRev.}A. Polkovnikov, K. Sengupta, A. Silva, and
M. Vengalattore, Rev. Mod. Phys. \textbf{83}, 863 (2011).

\bibitem{Higher-resolution1}W. S. Bakr, A. Peng, M. E. Tai, R. Ma,
J. Simon, J. Gillen, S. Foelling, L. Pollet, and M. Greiner, Science
\textbf{329}, 547 (2010).

\bibitem{Higher-resolution2}C.-L. Hung, X. Zhang, L.-C. Ha, S.-K.
Tung, N. Gemelke, and C. Chin, New. J. Phys. \textbf{13}, 075019 (2011).

\bibitem{ex4}P. Makotyn, C. E. Klauss, D. L. Goldberger, E. A. Cornell,
and D. S. Jin, Nature Phys. \textbf{9}, 512 (2013).

\bibitem{thermal1}M. Rigol, V. Dunjko, and M. Olshanii, Nature (London)
\textbf{452}, 854 (2008).

\bibitem{thermal2}J. Dziarmaga, Adv. Phys. \textbf{59}, 1063 (2010).

\bibitem{thermal3}T. Kinoshita, T. Wenger, and D. S. Weiss, Nature
(London) \textbf{440}, 900 (2006).

\bibitem{thermal4}R. Barnett, A. Polkovnikov, and M. Vengalattore,
Phys. Rev. A \textbf{84}, 023606 (2011).

\bibitem{thermal5}T. Kitagawa, A. Imambekov, J. Schmiedmayer, and
E. Demler, New J. Phys. \textbf{13}, 073018 (2011).

\bibitem{thermal6}M. Gring, M. Kuhnert, T. Langen, T. Kitagawa, B.
Rauer, M. Schreitl, I. Mazets, D. A. Smith, E. Demler, and J. Schmiedmayer,
Science \textbf{337}, 1318 (2012).

\bibitem{int1}M. Rigol, V. Dunjko, V. Yurovsky, and M. Olshanii,
Phys. Rev. Lett. \textbf{98}, 050405 (2007).

\bibitem{int2}M. A. Cazalilla, Phys. Rev. Lett. \textbf{97}, 156403
(2006).

\bibitem{int3}A. Iucci and M. A. Cazalilla, Phys. Rev. A \textbf{80},
063619 (2009).

\bibitem{int4}N. Nessi and A. Iucci, Phys. Rev. B \textbf{87}, 085137
(2013).

\bibitem{int5}M. A. Cazalilla, A. Iucci, and M. C. Chung, Phys. Rev.
E \textbf{85}, 011133 (2012).

\bibitem{phasetran1}Matthew S. Foster, Maxim Dzero, Victor Gurarie,
and Emil A. Yuzbashyan, Phys. Rev. B \textbf{88}, 104511 (2013).

\bibitem{phasetran2}Manuel Endres, Takeshi Fukuhara, David Pekker,
Marc Cheneau, Peter Schau{\ss}, Christian Gross, Eugene Demler,
Stefan Kuhr, and Immanuel Bloch, Nature\textbf{487}, 454 (2012).

\bibitem{phasetran3}Toshiya Kinoshita, Trevor Wenger, and David S.
Weiss, Nature \textbf{440}, 900 (2006).

\bibitem{tran4}L. E. Sadler, J. M. Higbie, S. R. Leslie, M. Vengalattore,
and D. M. Stamper-Kurn, Nature \textbf{443}, 312 (2006).

\bibitem{tran5}S. Hofferberth, I. Lesanovsky, B. Fischer, T. Schumm,
and J. Schmiedmayer, Nature \textbf{449}, 324 (2007).

\bibitem{tran6}Wojciech H. Zurek, Uwe Dorner, and Peter Zoller, Phys.
Rev. Lett. \textbf{95}, 105701 (2005).

\bibitem{changeEpara1}M. Greiner, O. Mandel, T. W. H?nsch, and I.
Bloch, Nature (London) \textbf{419}, 51 (2002).

\bibitem{changeEpara2Bragg}H. Miyake, G. A. Siviloglou, G.
Puentes, D. E. Pritchard, W. Ketterle, and D. M. Weld, Phys. Rev.
Lett. \textbf{107}, 175302 (2011).

\bibitem{changeEpara3insitu}S. Trotzky, Y-A. Chen, A. Flesch,
I. P. McCulloch, U. Schollw\"{o}ck, J. Eisert, and I. Bloch, Nat. Phys.
\textbf{8}, 325 (2012).

\bibitem{changeEpara4ToF}S. S. Natu, D. C. McKay, B. DeMarco,
and E. J. Mueller, Phys. Rev. A \textbf{85}, 061601(R) (2012).

\bibitem{changeEpara5insitu}M. Cheneau, P. Barmettler, D.
Poletti, M. Endres, P. Schaub, T. Fukuhara, C. Gross, I. Bloch, C.
Kollath. and S. Kuhr, Nature (London) \textbf{481}, 484 (2012).

\bibitem{changeEpara6}C.-L. Hung, V. Gurarie, and C. Chin, Science
\textbf{341}, 1213 (2013).

\bibitem{changeEpara7}A. Imambekov, I. E. Mazets, D. S. Petrov, V.
Gritsev, S. Manz, S. Hofferberth, T. Schumm, E. Demler, and J. Schmiedmayer,
Phys. Rev. A \textbf{80}, 033604 (2009).

\bibitem{Q1}S. S. Natu and E. J. Mueller, Phys. Rev. A \textbf{87},
053607 (2013).

\bibitem{Q2}S. S. Natu and E. J. Mueller, Phys. Rev. A \textbf{87},
063616 (2013).

\bibitem{Q3}Xiao Yin and Leo Radzihovsky, Phys. Rev. A \textbf{88},
063611 (2013).

\bibitem{LE1}Bal\'{a}zs D\'{o}ra, Frank Pollmann, J\'{o}zsef Fort\'{a}gh, and Gergely
Zar\'{a}nd, Phys. Rev. Lett. \textbf{111}, 046402 (2013).

\bibitem{LE2}Rashi Sachdeva, Tanay Nag, Amit Agarwal, and Amit Dutta,
Phys. Rev. B \textbf{90}, 045421 (2014).

\bibitem{Contact1}S. Tan, Ann. Phys. (NY) \textbf{323}, 2952 (2008);
\textbf{323}, 2971 (2008); \textbf{323}, 2987 (2008).

\bibitem{Contact5}E. Braaten and L. Platter, Phys. Rev. Lett. \textbf{100},
205301 (2008).

\bibitem{Contact6}F. Werner, L. Tarruell, and Y. Castin, Eur. Phys.
J. B \textbf{68}, 401 (2009).

\bibitem{Contact7}S. Zhang and A. J. Leggett, Phys. Rev. A \textbf{79},
023601 (2009).

\bibitem{Contact2Bragg}E. D. Kuhnle, H. Hu, X. J. Liu, P.
Dyke, M. Mark, P. D. Drummond, P. Hannaford, and C. J. Vale, Phys.
Rev. Lett. \textbf{105}, 070402 (2010).

\bibitem{Contact3}J.M.Diederix, T. C. F. vanHeijst, and H. T. C.
Stoof, Phys. Rev A \textbf{84}, 033618 (2011).

\bibitem{Contact4}R. J. Wild, P. Makotyn, J. M. Pino, E. A. Cornell,
and D. S. Jin. Phys. Rev. Lett. \textbf{108}, 145305 (2012).

\bibitem{Bragg}T. A. Corcovilos, S. K. Baur, J.M. Hitchcock, E. J.
Mueller, and R. G. Hulet, Phys. Rev. A \textbf{81}, 013415 (2010).

\bibitem{NoiseCorrelation1}E. Altman, E. Demler, and M. D. Lukin,
Phys. Rev A\textbf{ 70}, 013603 (2004).

\bibitem{NoiseCorrelation2}M. Greiner, C. A. Regal, J. T. Stewart,
and D. S. Jin, Phys. Rev. Lett. \textbf{94}, 110401 (2005).

\bibitem{n0-nNice}D. S. Petrov, G. V. Shlyapnikov, and J. T. M. Walraven,
Phys. Rev. Lett. \textbf{85}, 3745 (2000).

\end{thebibliography}
\end{document}